%% file: main.tex
\documentclass[conference]{IEEEtran}
\usepackage{fancyhdr}
\usepackage{cite}
\usepackage{amsmath,amssymb,amsfonts}
\usepackage{graphicx}
\usepackage{textcomp}
\usepackage{xcolor}
\usepackage{tcolorbox}
\usepackage{quantikz}
\usepackage{soul}
\usepackage{xspace}
\usepackage{float}
\usepackage{multirow}
\usepackage{grffile}
\usepackage{makecell}
\usepackage{algorithm}
\usepackage{algpseudocode}
\usepackage{comment}
\usepackage{url}
\usepackage[hidelinks]{hyperref}
\usepackage{booktabs}
\usepackage{enumitem}

\newcommand{\sol}{OptiPESS}

\begin{document}

\title{A GPU-Accelerated and Geometry-Aware Quantum-Classical Framework for Optimized PES Scans with CUDA-Q}

\author{%
\IEEEauthorblockN{Panayiotis Christou, Shuwen Kan, and Ying Mao}
\IEEEauthorblockA{\textit{Fordham University}\\
New York City, New York, USA\\
\{pc33, sk107, ymao41\}@fordham.edu}}

\maketitle

\thispagestyle{fancy}
\lhead{}
\rhead{}
\chead{}
\lfoot{
}
\rfoot{}
\cfoot{}
\renewcommand{\headrulewidth}{0pt}
\renewcommand{\footrulewidth}{0pt}

\begin{abstract}
Hybrid quantum-classical workloads, particularly Potential Energy Surface (PES) scans in quantum chemistry using batched VQE and QAOA, involve many structurally similar Hamiltonians differing only in coefficients. Existing warm-start and meta-optimization strategies reuse parameters between instances but still consume fixed quantum-classical resources, ignoring cross-instance structural similarities. We present OptiPESS, a GPU-accelerated runtime system for adaptive resource allocation in hybrid quantum-classical execution. OptiPESS introduces the Locality-Occupancy-Variance (LOV) metric, a lightweight, geometry-aware measure of Hamiltonian similarity, coupled with efficient alignment and Pauli-classification passes. Using LOV, OptiPESS performs distance-aware scheduling of iteration and shot budgets, predicting which instances can be skipped, quick-solved, or fully optimized. Implemented atop CUDA-Q with persistent multi-GPU execution and near-linear scaling, OptiPESS reduces classical and quantum evaluations by up to 78.3\% and wall-time by up to 74.8\% while maintaining baseline accuracy. This demonstrates how lightweight static analysis and adaptive scheduling can transform quantum-classical co-execution from fixed-budget loops into a predictive, resource-aware runtime.

\end{abstract}


\input{sections/introduction-new}
\input{sections/background-motivation}
\input{sections/system-architecture}

\input{sections/evaluation-new}
\input{sections/discussion-future-work}
\input{sections/conclusion-new}

\section*{Acknowledgments}
OpenAI ChatGPT~\cite{openai_chatgpt} and OpenAI Codex~\cite{openai_codex} were used in the Abstract and Sections~I--VI for language editing and refinement, including grammar correction and improvements to clarity and readability, and to assist in drafting the layout of Figure~1. No AI system was used to generate the research ideas, methods, experimental design, results, or analysis presented in this work. The authors reviewed, verified, and revised all AI-assisted content and remain fully responsible for the final manuscript. This research was supported in part by the National Science Foundation (NSF) under grant agreements 2329020, 2343535 and 2408229.

\IEEEtriggeratref{36}
\bibliographystyle{IEEEtran}
\bibliography{sections/references}

\end{document}

%% file: sections/introduction-new.tex
\section{Introduction}
\label{sec:intro}

Quantum computing offers immense potential across diverse fields~\cite{hatay2025top, golec2025quantum}, such as machine learning \cite{Wang_2024}, security \cite{Faruk_2022}, and scientific computing \cite{dalzell2023quantumalgorithmssurveyapplications}. Among these, Hamiltonian-based quantum algorithms form the foundation for modeling a wide range of physical, chemical, and combinatorial systems, and are therefore central to many practical quantum applications. However, realizing this potential requires overcoming significant engineering challenges in building stable, large-scale quantum machines \cite{ayyamgari_2023}. To address this, hybrid quantum-classical systems \cite{campos2024hybridquantumclassicalalgorithms}, which combine classical computing resources like GPUs with quantum processors, have emerged as a promising path forward. 

Classical high-performance computing resources are especially important in this setting because they can handle much of the numerical and coordination work around quantum execution. GPUs and HPC systems can speed up tasks such as Hamiltonian construction, parameter handling, batching, scheduling, and simulation, making them essential for fast and reliable operation of quantum processors in hybrid workflows. At the same time, their growing role highlights a basic reality of near-term quantum computing: performance depends not only on the quantum hardware itself, but also on how efficiently the surrounding classical system supports repeated quantum calls. This matters even more for variational methods, where constant back-and-forth between classical and quantum computation can quickly become a major source of overhead.

The importance of classical infrastructure is evident across multiple layers of the quantum computing stack. For example, quantum machine learning combines quantum processing with classical training and data handling~\cite{stein2022quclassi, cerezo2022challenges, l2024quantum, rodriguez2025survey}, while hardware-aware quantum compilers address mapping, routing, and scheduling~\cite{bahrani2026resource, sharma2026space, du2026dismap, du2026efficient, palma2025hardware} under device-specific constraints~\cite{kan2024scalable, kan2025circuit}. Quantum error correction (QEC), a foundation of fault-tolerant quantum computing (FTQC), further requires tightly coordinated syndrome measurements, real-time classical decoding, and conditional quantum operations~\cite{kan2025sparo, christou2026hybridqc}.

Due to the advent of hybrid quantum-classical systems, hybrid quantum–classical algorithms have become the dominant execution model for near-term quantum computing. In the Variational Quantum Eigensolver (VQE)\cite{Tilly_2022}, a quantum processor evaluates parameterized circuits while a classical optimizer updates the parameters to minimize the expected energy. This iterative pattern allows noisy intermediate-scale quantum (NISQ) devices\cite{campos2024hybridquantumclassicalalgorithms} to tackle useful problems, but it also introduces substantial overhead: every iteration incurs host–device communication, repeated circuit execution, and a fixed amount of classical optimization regardless of how similar consecutive problem instances are~\cite{Carreras:2025rro,Scriva_2024}.

An important setting is \emph{batched Hamiltonian workloads}, where many closely related Hamiltonians must be solved. A canonical example is a Potential Energy Surface (PES) scan in quantum chemistry~\cite{hu2022benchmarkingvariationalquantumeigensolvers}, where each molecular geometry defines a slightly different Hamiltonian but the operator structure is largely shared. Existing accelerations such as parameter transfer and warm-starting~\cite{Skogh_2023,meyer2024warm} reuse optimized parameters between geometries, yet they still run full VQE loops—with the same maximum iterations and shot counts—for every geometry. As a result, they ignore the smooth “geometry” of the workload in coefficient space and spend unnecessary computational effort on easy, low-variation instances.

We argue that, in a warm-start setting, the structural similarity between Hamiltonians provides sufficient information to predict the quantum and classical resources a VQE instance will require. In this work we present \sol, a geometry-aware runtime system that uses lightweight static analysis of batched Hamiltonians to assign instance-specific budgets \emph{before} executing quantum circuits. \sol~introduces a Locality–Occupancy–Variance (LOV) metric that operates directly on aligned Pauli coefficients, emphasizing operator classes that are structurally important, frequently active, and sensitive to changes in molecular geometry. From these statistics, \sol~constructs a similarity profile over the PES scan and converts it into per-geometry iteration and shot budgets via distance-aware decision bands (skip/quick/full).

At execution time, \sol~runs a persistent CUDA-Q VQE solver that keeps the ansatz, statevector, and optimizer state resident on GPUs. Consecutive geometries are warm-started from the previous solution, gradients are computed via distributed parameter-shift across multiple GPUs, and the precomputed budgets determine how aggressively each geometry is optimized. Easy, high-similarity geometries are either skipped or solved with few iterations and low shot counts. Only high-variation geometries pay the full optimization cost.

On a suite of PES benchmarks with up to 1{,}600 Pauli terms, \sol~reduces total evaluations and wall-clock simulation time by up to 78\% and achieves up to $3.6\times$ end-to-end speedup over strong baselines, while preserving chemical accuracy. Although our evaluation focuses on PES scans, the same mechanism applies to other batched Hamiltonian workloads where instances share structure but differ in coefficients. Our contributions are as follows:
\begin{itemize}
\item We present \sol, a runtime system that turns fixed-budget VQE PES scans into an adaptive, geometry-aware execution by predicting instance difficulty from Hamiltonian statistics.
\item We introduce the LOV metric, a GPU-friendly Hamiltonian similarity measure that combines locality, occupancy, and variance at the Pauli-class level to drive budget assignment.
\item We implement a multi-GPU CUDA-Q solver with persistent state, optimizer-state transfer, and distributed gradient evaluation that consumes LOV-derived budgets.
\item Evaluated with realistic quantum-chemistry PES workloads, \sol~ reduces classical and quantum evaluations by up to 78.3\% and wall-time
by up to 74.8\% while maintaining baseline accuracy.

\end{itemize}

%% file: sections/background-motivation.tex
\section{Background and Related Work}
\label{sec:background}

\subsection{Hybrid Quantum--Classical Execution}

Hybrid quantum--classical algorithms alternate between quantum circuit evaluation and classical optimization. In each iteration, a quantum processor prepares a parameterized state $|\psi(\boldsymbol{\theta})\rangle$, measures the expected energy $\langle \psi|H|\psi\rangle$, and a classical optimizer updates the parameters. This structure defines the Variational Quantum Eigensolver (VQE)~\cite{Peruzzo_2014,Tilly_2022}, one of the most widely adopted algorithms on NISQ devices~\cite{campos2024hybridquantumclassicalalgorithms}.

Although easy to implement, these workloads are expensive to execute in practice: every optimization step requires thousands of circuit evaluations (shots), frequent CPU--QPU synchronization, and repeated compilation or kernel initialization~\cite{Scriva_2024}. Runtime grows with both the number of parameters and the number of required optimization steps, creating significant overhead in applications where many closely related Hamiltonians must be solved (See Sec. \ref{subsec:batched_hams}).

\subsection{Hamiltonians and Pauli Representations}
\label{subsec:ham_pauli_str}

Many quantum applications are governed by a problem Hamiltonian $H$, a Hermitian operator whose eigenvalues encode the objective (e.g., molecular energies or cost functions). On quantum hardware, $H$ is represented through its Pauli decomposition:
$$
  H = \sum_{i=1}^{P} c_i P_i,\quad
  P_i \in \{I,X,Y,Z\}^{\otimes n},
$$
where $c_i\in\mathbb{R}$ are Pauli coefficients. 

In electronic structure, $H$ originates from a fermionic Hamiltonian with one- and two-electron integrals~\cite{naeij2024molecularelectronicstructurecalculation}, which is mapped to qubits through Jordan--Wigner~\cite{Li_2022}, Bravyi--Kitaev~\cite{Seeley_2012}, or related encodings~\cite{Fellner_2022,Jiang_2020}. After encoding, the full problem is described by its coefficient vector $\boldsymbol{c}=(c_1,\dots,c_P)^\top$~\cite{zhu2025quantumcompilerdesignqubit}.

A key property exploited in this work is that, across many practical workloads (e.g., potential energy surface scans), the \emph{Pauli structure} remains stable while the coefficients vary smoothly. This makes the coefficient vectors an appropriate representation for computing inter-instance similarity, provided the Pauli basis is aligned across Hamiltonians (See Sec. \ref{subsec:ham_align}). This observation underpins the design of \sol.

\subsection{Batched Hamiltonian Workloads}
\label{subsec:batched_hams}
In quantum chemistry, a potential energy surface (PES) scan evaluates a sequence of molecular geometries. The resulting Hamiltonians $H_0,\dots,H_{N-1}$ differ primarily in their coefficients~\cite{hu2022benchmarkingvariationalquantumeigensolvers}. Similar batched structures arise in parameter sweeps and other iterative modeling tasks.

Despite the smooth evolution of $H_g$ with geometry, standard VQE pipelines treat every Hamiltonian as an independent optimization problem, repeating compilation, warm-start initialization, and full-iteration optimization loops from scratch. This results in redundant quantum evaluations and unnecessary classical optimization (See Sec. \ref{subsec:limits_static_resources}).

\subsection{Warm-Start and Parameter-Transfer Methods}
\label{subsec:warm_start}

Several techniques attempt to exploit cross-instance similarity. Parameter transfer and warm-start strategies reuse optimized parameters from a nearby instance to accelerate convergence~\cite{Skogh_2023,meyer2024warm}. Classical preconditioners, adaptive initialization, and statevector reuse have also been proposed. While these methods reduce the number of “wasted” early iterations, they still execute full VQE loops for every Hamiltonian and do not determine when running fewer iterations or shots is sufficient. Critically, they lack a measure of similarity that is robust, physically meaningful (See Sec. \ref{subsec:limits_similarity_metrics}), and cheap to compute across the entire workload.

\subsection{Limitations of Existing Similarity Metrics}
\label{subsec:limits_similarity_metrics}
Prior similarity measures compare Hamiltonians or states pairwise using Frobenius norms and Hilbert--Schmidt distances~\cite{Yao_2016,kuzmak2021measuringdistancequantumstates,jahromi2020hilbertschmidtspeedefficienttool}. They do not capture how coefficient perturbations influence optimization difficulty or energy landscapes and suffer from two issues:

\begin{itemize}
\item \emph{Representation instability.} Orbital rotations, tapering, or mapping choices may change coefficients without altering the underlying physics~\cite{boys1960construction,pulay1971abinitio,bravyi2017tapering}.
\item \emph{Lack of workload context.} Pairwise distances ignore cross-instance structure essential for determining which Hamiltonians are “easy” or “hard” within a batch.
\end{itemize}

Thus, existing metrics offer limited utility for predictive scheduling.

\subsection{Limitations of Static Resource Allocation}
\label{subsec:limits_static_resources}
Standard hybrid runtimes (Qiskit Runtime, CUDA-Q, tket) assign fixed iteration and shot budgets regardless of Hamiltonian similarity or convergence progress. This behavior is appropriate for isolated VQE runs, but inefficient for batched workloads where many instances are nearly identical and do not require full optimization effort. Furthermore, shot allocation remains largely static. Excessive shots add noise-averaging overhead, while insufficient shots hinder convergence. No current system allocates resources adaptively based on Hamiltonian similarity across a workload.

\subsection{Motivation}

These limitations motivate a geometry-aware approach that reasons directly over the Pauli-coefficient evolution of the workload. If two Hamiltonians exhibit minimal coefficient change, their optimal parameters and energies are expected to be similar, while large coefficient variation indicates increased optimization difficulty. Exploiting this structure allows a runtime to \emph{predict} which instances require full optimization and which can be skipped or solved with reduced budgets.

\sol~builds on this insight, introducing a low-cost Hamiltonian similarity model and an adaptive scheduler that transforms fixed-budget hybrid execution into a workload-aware, resource-efficient runtime. 

%% file: sections/system-architecture.tex
\section{System Architecture}
\label{sec:system_architecture}

\begin{figure*}[t!]
    \centering
    \includegraphics[width=\textwidth]{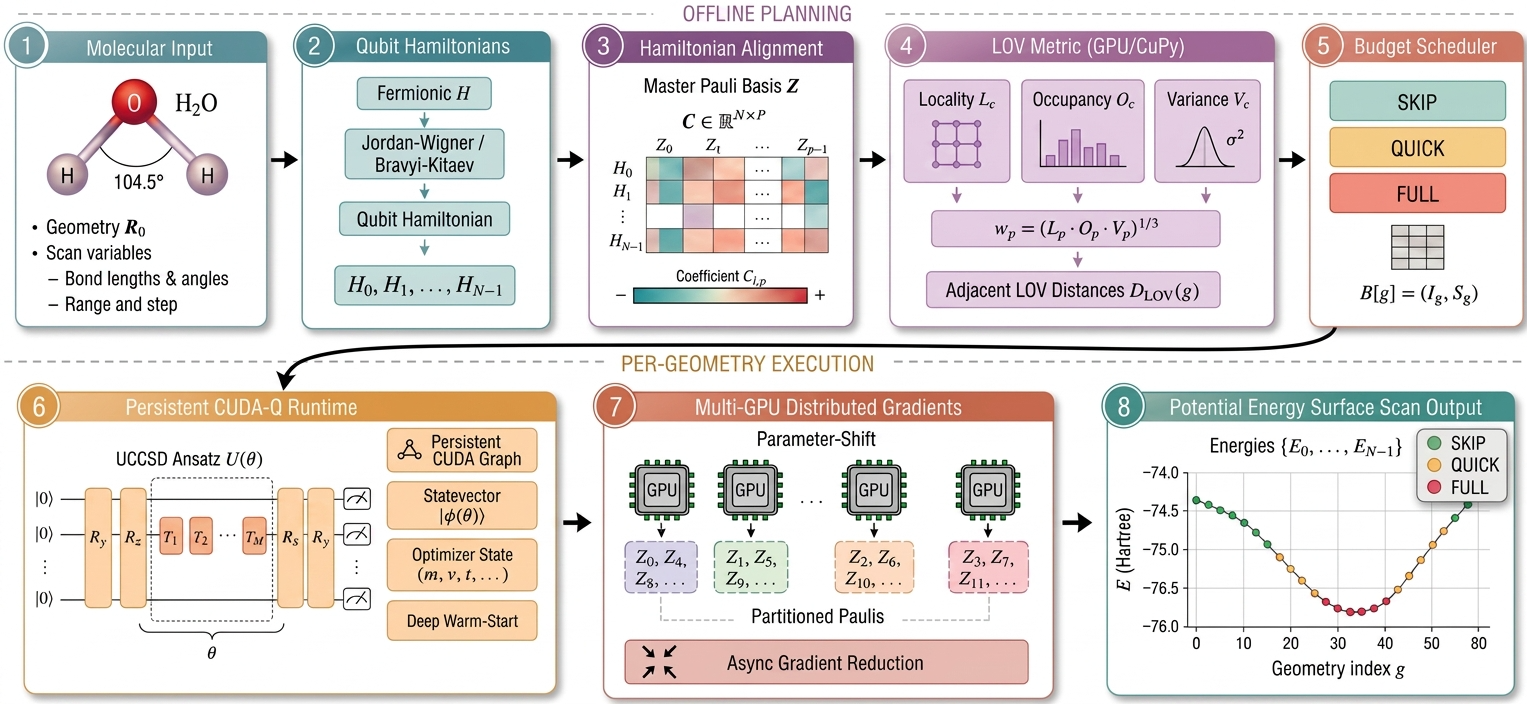}
    \caption{\sol's end-to-end PES/VQE pipeline. Offline planning maps each molecular geometry to a qubit Hamiltonian, aligns the Hamiltonians, computes LOV distances with CuPy, and assigns skip, quick, or full budgets. Per-geometry execution reuses a persistent CUDA-Q UCCSD runtime and distributes parameter-shift gradients over the available GPUs before producing the PES.}
    \label{fig:high_level_system_diag}
\end{figure*}

\subsection{Overview and Design Goals}

Fig.~\ref{fig:high_level_system_diag} shows the \sol~pipeline from molecular input to adaptive PES evaluation. First, the user specifies an initial molecular geometry, scan variables (e.g., bond lengths and angles), and the scan range and step (panel 1). For each geometry, the chemistry preprocessing stack maps the fermionic molecular Hamiltonian to a qubit Hamiltonian (panel 2), yielding a sequence $\{H_0,\dots,H_{N-1}\}$.

Then \sol~collects these Hamiltonians into one aligned Pauli matrix by padding all coefficient vectors into a common Pauli basis (panel 3 and Sec.~\ref{subsec:ham_align}). Offline similarity analysis computes the locality, occupancy, and variability (LOV) statistics and adjacent-geometry distances (panel 4 and Sec.~\ref{subsec:lov_distance}). The distance-aware scheduler converts these distances into per-geometry iteration and shot budgets (panel 5 and Sec.~\ref{subsec:distance_aware_scheduler_compact}). Panels 6 and 7 show the persistent runtime and multi-GPU gradient evaluation, while panel 8 records PES energies and scheduling regimes (Sec.~\ref{subsec:multi_gpu_runtime}).

\sol~is designed around three system goals: (i) predict resource requirements per geometry from lightweight Hamiltonian statistics, (ii) maximize cross-geometry reuse of parameters, optimizer state, and measurement effort, and (iii) scale efficiently on GPU-backed simulators and, in principle, hardware-attached runtimes. The remainder of this section describes GPU-accelerated similarity analysis (Sec.~\ref{subsec:gpu_similarity}), distance-aware budget scheduling (Sec.~\ref{subsec:distance_aware_scheduler_compact}), and the hybrid quantum-classical runtime (Sec.~\ref{subsec:multi_gpu_runtime}), followed by the end-to-end PES workflow (Sec.~\ref{subsec:pipeline}).

\subsection{GPU-Accelerated Similarity Analysis}
\label{subsec:gpu_similarity}

The middle portion of Fig.~\ref{fig:high_level_system_diag} encodes the offline analysis that turns a set of Hamiltonians into a compact similarity structure. As discussed in Sec.~\ref{subsec:ham_pauli_str}, each Hamiltonian $H_g$ is represented by Pauli coefficients. \sol~first aligns these Hamiltonians into a common basis (Sec \ref{subsec:ham_align}) and then computes class-informed LOV weights (Sec \ref{subsec:lov_distance}) to emphasize physically meaningful variations.

Let $\mathbf{C} \in \mathbb{R}^{N \times P}$ denote the aligned coefficient matrix over $N$ geometries and $P$ unique Pauli terms. Using CuPy on GPU, each Pauli string $p$ receives a weight
\begin{equation}
  w_p = (L_p O_p V_p)^{1/3}, \qquad
  W = \mathrm{diag}(w_1,\dots,w_P),
  \label{eq:love_weight}
\end{equation}
where $L_p$, $O_p$, and $V_p$ are derived from the locality, occupancy, and variability of the Pauli class $c_p$ to which $p$ belongs.

Pauli strings are grouped into chemically motivated classes such as one-body ($Z_i$), two-body Coulomb ($Z_iZ_j$), spin-flip ($X/Y$), and mixed-parity or non-local couplings (following standard chemical partitioning conventions \cite{ayyamgari_2023, pulay1971abinitio, Niklasson_2004, bravyi2018advantage, aszabo82qchem, olsen_bible}. For each class $c$:
\begin{equation}
  L_{c} = 2^{-k_{c}},
  \quad
  k_{c} = \min_{P_{i}\in c} \kappa_{i},
  \label{eq:loc_def}
\end{equation}
\begin{equation}
  O_{c} = \frac{n_{c}}{\max_{d} n_{d}},\qquad
  n_{c} = \bigl|\{\, g : \exists\, P_{i}\in c,\; c_{i}(g)\neq 0 \}\bigr|,
  \label{eq:occ_def}
\end{equation}
\begin{equation}
  V_{c} = \frac{\sigma_{c}}{\max_{d}\sigma_{d}},\qquad
  \sigma_{c} = \sqrt{\frac{1}{n_{c} N_{c}} \sum_{g\in\mathcal{G}_{c}} \sum_{P_{i}\in c} \bigl(c_{i}(g)-\bar c_{c}\bigr)^{2}}.
  \label{eq:var_def}
\end{equation}
Each Pauli string inherits $w_p = (L_{c_p} O_{c_p} V_{c_p})^{1/3}$, emphasizing classes that are low-body, frequently occupied, and coefficient-sensitive across the scan. Panel 4 of Fig.~\ref{fig:high_level_system_diag} summarizes these statistics and their projection into adjacent LOV distances.

\paragraph*{Hamiltonian Alignment.}
\label{subsec:ham_align}
The \emph{Hamiltonian Alignment} panel corresponds to the alignment step that enables GPU processing in \sol. Each $H_g$ may have a slightly different set of Pauli strings. \sol~constructs a unified basis $\mathcal{Z}$ and pads missing coefficients with zeros to obtain $\mathbf{C}$ with consistent columns. It initializes a master basis and adds every Pauli string not already present while traversing the Hamiltonians.


\paragraph*{LOV Embedding and Distances.}
\label{subsec:lov_distance}
Using the weights $W$, the aligned Hamiltonians are embedded into
$
  \mathbf{X} = \mathbf{C} W,
$
and adjacent LOV distances are computed on GPU (See Sec. \ref{subsubsec:similarity_analysis}).
From these we extract the maximum scan variation $D_{\max}$ (Sec.~\ref{subsubsec:similarity_analysis}), which summarizes PES heterogeneity and provides a global scale for identifying high-similarity zones (Sec.~\ref{subsubsec:distance_aware_scheduler}). Panel 5 of Fig.~\ref{fig:high_level_system_diag} shows how these zones determine budget allocation.

\subsection{Distance-Aware Budget Scheduling}
\label{subsec:distance_aware_scheduler_compact}

The LOV-guided scheduler converts distances $\{D_{\mathrm{LOV}}(g)\}$ into per-geometry iteration and shot budgets $(I_g,S_g)$, so the runtime knows in advance which geometries to skip, solve approximately, or fully optimize (See Sec. \ref{subsubsec:distance_aware_scheduler}).


This offline pass produces a compact budget table $\mathcal{B}[g]=(I_g,S_g)$ that the runtime consults during execution (Sec.~\ref{subsubsec:multi_gpu_vqe}). Panel 5 of Fig.~\ref{fig:high_level_system_diag} corresponds to this banding and budget assignment.

\subsection{Hybrid Quantum--Classical Runtime}
\label{subsec:multi_gpu_runtime}

Panels 6 and 7 of Fig.~\ref{fig:high_level_system_diag} represent the \sol~runtime, which executes the PES scan on a CUDA-Q backend using the precomputed budgets. The runtime keeps the UCCSD ansatz compilation persistent, carries optimizer state across geometries, and parallelizes gradient evaluation across the available GPUs (Sec.~\ref{subsubsec:multi_gpu_vqe}). Panel 8 shows the resulting PES and each geometry's execution regime.

\paragraph*{Persistent Solver and Deep Warm-Start.}
\sol~initializes a long-lived CUDA-Q context and compiles the variational ansatz $U(\theta)$ into a persistent CUDA graph once per workload. Geometry $R_0$ is solved from a Hartree--Fock initialization~\cite{hf_2020}, yielding $(\theta_0,E_0,\text{opt\_state}_0)$. For $g>0$, the solver warm-starts from \\
$(\theta_{g-1},\text{opt\_state}_{g-1})$ and configures its iteration and shot budgets from $\mathcal{B}[g]$. Skip-band geometries bypass VQE entirely by copying $(\theta_{g-1},E_{g-1})$. Quick-band geometries run fewer optimizer steps and shots, while full-band geometries use the maximum budgets.

\paragraph*{Multi-GPU Gradient Evaluation.}
Within each VQE call, \textit{analytic parameter-shift gradients are distributed across GPUs}. Pauli terms are partitioned among devices, each device accumulates partial gradients locally, and a reduction on a primary device updates $\theta$ in place. 

\begin{algorithm}[h]
\caption{\sol~End-to-End Execution}
\label{algo:optipess_framework}
\begin{algorithmic}[1]
\Require Hamiltonians $\{H_0,\dots,H_{N-1}\}$, ansatz $U(\theta)$, max budgets $(I_{\max}, S_{\max})$, fractions $(\eta_{\text{skip}}, \eta_{\text{quick}})$
\Statex \textbf{// Offline Similarity Analysis}
\State $\mathbf{C} \gets$ Align $\{H_0,\dots,H_{N-1}\}$  and construct aligned matrix
\State Classify Pauli strings; compute $(L_c,O_c,V_c)$ and per-string $w_p$
\State $\mathbf{X} \gets \mathbf{C} W$
\For{$g = 1$ to $N-1$} \State $D_{\mathrm{LOV}}(g) \gets \|x_g - x_{g-1}\|_2$ \EndFor
\State Compute $D_{\max}$; set $\varepsilon_{\text{skip}} \gets \eta_{\text{skip}} D_{\max}$, $\varepsilon_{\text{quick}} \gets \eta_{\text{quick}} D_{\max}$
\Statex \textbf{// Budget-Aware Scheduling}
\For{$g = 0$ to $N-1$}
  \If{$g = 0$}
    \State $I_g,S_g \gets I_{\max},S_{\max}$ \Comment{Bootstrap}
  \ElsIf{$D_{\mathrm{LOV}}(g) \le \varepsilon_{\text{skip}}$}
    \State $I_g,S_g \gets 0,0$ \Comment{Skip band}
  \ElsIf{$D_{\mathrm{LOV}}(g) \le \varepsilon_{\text{quick}}$}
    \State $\alpha \gets 1 - D_{\mathrm{LOV}}(g) / D_{\max}$; 
           $c_I \gets \lfloor \eta_{\text{quick}} I_{\max} \rfloor$; 
           $c_S \gets \lfloor \eta_{\text{quick}} S_{\max} \rfloor$
    \State $I_g \gets \max(1, \lfloor \alpha c_I \rfloor)$; 
           $S_g \gets \max(1, \lfloor \alpha c_S \rfloor)$
  \Else
    \State $I_g,S_g \gets I_{\max},S_{\max}$ \Comment{Full band}
  \EndIf
\EndFor
\State Store budget table $\mathcal{B}[g] = (I_g,S_g)$
\Statex \textbf{// CUDA-Q Runtime}
\State Initialize CUDA-Q solver and persistent CUDA graph for $U(\theta)$
\State Solve $H_0$ with HF init; cache $(\theta_0, E_0, \text{opt\_state}_0)$
\For{$g = 1$ to $N-1$}
  \State Read $(I_g,S_g)$ from $\mathcal{B}[g]$
  \If{$I_g = 0$ \textbf{and} $S_g = 0$}
    \State $(\theta_g,E_g) \gets (\theta_{g-1},E_{g-1})$; \textbf{continue}
  \EndIf
  \State Warm-start solver with $(\theta_{g-1}, \text{opt\_state}_{g-1})$; configure $(I_g,S_g)$
  \State Run VQE on CUDA-Q; obtain $(\theta_g,E_g)$ and update $\text{opt\_state}_g$
\EndFor
\State \Return Energies $\{E_0,\dots,E_{N-1}\}$
\end{algorithmic}
\end{algorithm}

\subsection{LOV Cost and Memory Analysis}
\label{subsec:lov_cost_analysis}

The LOV analysis is a one-time preprocessing cost over the aligned workload, whereas Plain VQE repeatedly pays the optimization cost at every geometry. For $G$ geometries and $P$ aligned Pauli terms, the static cost decomposes as
\begin{equation}
\begin{aligned}
T_{\mathrm{static}}
  &=T_{\mathrm{align}}+T_{\mathrm{weights}}+T_{\mathrm{dist}}+T_{\mathrm{budget}}\\
  &=O\!\left(\sum_g P_g+GP\right).
\end{aligned}
\label{eq:static_runtime}
\end{equation}
The corresponding end-to-end costs are
\begin{equation}
\begin{aligned}
T_{\mathrm{OptiPESS}}&=T_{\mathrm{static}}+\sum_g I_g C_g(S_g),\\
T_{\mathrm{Plain}}&\approx\sum_g I_{\max}C_g(S_{\max}),
\end{aligned}
\label{eq:end_to_end_runtime}
\end{equation}
where $C_g(S_g)$ is the cost of an optimizer step or expectation group. Skip points have $I_g=S_g=0$, while quick points use reduced budgets. The static phase is therefore amortized by the repeated VQE work avoided across the scan.

The LOV metric computes a class-weighted $\ell_2$ difference between Hamiltonian coefficients with
$\mathrm{Mem}_{\mathrm{LOV}}\!\approx\!P(16+\tfrac{n}{2})$ bytes, where $P\!\ll\!4^n$ is the number of Pauli terms after tapering. Its storage scales with the sparse Pauli representation rather than the full operator space and remains only a few MB for the benchmark sizes considered here.

\subsection{\sol~ Quantum-Classical Pipeline}
\label{subsec:pipeline}

\paragraph*{End-to-End Execution.}
Algorithm~\ref{algo:optipess_framework} summarizes the full control flow: During \emph{Offline Similarity Analysis} we build and align Hamiltonians (Lines 1-2) and compute LOV embeddings and distances (Lines 3-6). Then the \emph{Distance Aware Budget Scheduler} assigns budgets (Lines 7-20) and the \emph{Hybrid Runtime} executes a band-aware, warm-started VQE loop over all geometries by consuming precomputed budgets (Lines 21-31).

\subsubsection{Offline Similarity Analysis}
\label{subsubsec:similarity_analysis}
During Offline Similarity Analysis all static data required for adaptive scheduling of the VQE scans is precomputed. The inputs consist of a molecule \(\mathcal{M}\) and an ordered grid of geometries \(\{R_0, R_1, \dots, R_{N-1}\}\). For each geometry \(R_g\), we generate its qubit Hamiltonian via quantum chemistry preprocessing (Sec.~\ref{subsec:ham_pauli_str}). This data is processed using Hamiltonian Alignment (Alg.~\ref{algo:align_hamiltonians}) and Classify Paulis (Alg.~\ref{algo:classify_paulis}) to compute LOV weights, distances, the maximum global deviation, and adaptive workload thresholds on GPU. The distance-aware scheduler (Sec.~\ref{subsubsec:distance_aware_scheduler}) consumes these values to assign each geometry's regime and resource budget.

Two very important algorithms that enable the GPU acceleration of the LOV metric are Hamiltonian Alignment (Algorithm \ref{algo:align_hamiltonians}) and ClassifyPaulis (Algorithm \ref{algo:classify_paulis}). These sub-routines inside Algorithm \ref{algo:optipess_framework} are essential in creating the matrix form of the Hamiltonians on which the GPU operates on to efficiently calculate the LOV similarity distances with minimal data movement and high speed. We explain both sub-routines below.

\paragraph{Hamiltonian Alignment.} Each geometry’s Hamiltonian \(H_g\) contains a different set of Pauli terms
(Sec.~\ref{subsec:ham_pauli_str}).
To enable vectorized similarity computation, we perform
\emph{Hamiltonian alignment} by constructing a global Pauli basis:
starting from geometry \(R_0\), we iterate across all geometries, \(N\), and append any Pauli strings not encountered yet to a master list. \(P\) is the final length of the master list after iterating across all Paulis across all geometries is completed.
Each \(H_g\) is expanded into a coefficient vector
\(c^{(g)}\!\in\!\mathbb{R}^P\), padding zeros up to length \(P\) for missing terms to ensure all of the Hamiltonians are of equal length.
Stacking these vectors yields the aligned coefficient matrix
\(\mathbf{C}\in\mathbb{R}^{N\times P}\),
where each column corresponds to a unique Pauli string shared across all geometries. Algorithm ~\ref{algo:align_hamiltonians} shows the iterative process for Hamiltonian Alignment.

\begin{algorithm}[h]
\caption{\textsc{Align Hamiltonians}}
\label{algo:align_hamiltonians}
\begin{algorithmic}[1]
\Require set of qubit Hamiltonians
        $\mathcal{H}=\{H_Q^{(g)}\}_{g=0}^{G-1}$,\;
        each $H_Q^{(g)}=\sum_j c^{(g)}_j S^{(g)}_j$
\Ensure  master label list $\mathcal{L}$ and
        coefficient matrix $\mathbf{C}\!\in\!\mathbb{R}^{G\times P}$
\Statex{\textbf{Stage 1 — Build master label list}}
\State $\mathcal{L}\leftarrow[\,]$ \Comment{ordered list}
\State \texttt{index}$\!\leftarrow$ empty hash\,map
\For{$g\gets0$ \textbf{to} $G-1$}
      \For{$S\in H_Q^{(g)}$}
            \If{$S$ \textbf{not in} \texttt{index}}
                  \State append $S$ to $\mathcal{L}$
                  \State \texttt{index[$S$]} $\leftarrow$
                         $\lvert\mathcal{L}\rvert-1$
            \EndIf
      \EndFor
\EndFor
\State $P \leftarrow \lvert\mathcal{L}\rvert$
\Statex{\textbf{Stage 2 — Populate coefficient matrix}}
\State $\mathbf{C}\leftarrow \mathbf{0}_{G\times P}$
\For{$g\gets0$ \textbf{to} $G-1$}
      \For{$(c,S)\in H_Q^{(g)}$}
            \State $j \leftarrow$ \texttt{index[$S$]}
            \State $\mathbf{C}_{g,j} \leftarrow c$
      \EndFor
\EndFor
\State \Return $(\mathcal{L},\mathbf{C})$
\end{algorithmic}
\end{algorithm}

\paragraph{Classifying Pauli Operators.} Aligned Pauli strings are grouped into structural classes as explained in Section \ref{subsec:gpu_similarity} to enable class-weighted LOV computation. Each Pauli string is assigned a class identifier according to rule-based pattern matching on its non-identity composition i.e. how many and which non-identity single-qubit Pauli operators $(X,Y,Z)$ make up the string. This decomposition dictates the structural class of that Pauli string forming a class table used to compute class-specific LOV weights. It is important to note that since the Hamiltonians are aligned and stacked as a matrix, every column contains Paulis of the same structural class that differ only by their coefficient. This enables fast calculation of LOV weights through efficient columnwise operations. The exact algorithmic formulation of the ClassifyPaulis sub-routine is given in Algorithm \ref{algo:classify_paulis}.

\begin{algorithm}[h]
\caption{\textsc{Classify Paulis}}
\label{algo:classify_paulis}
\begin{algorithmic}[1]
\Require Pauli list $\mathcal{P}=\{P_1,\dots,P_M\}$;\\
        class table
        $\{\mathcal{C}_k\}_{k=0}^{C-1}$
        with boolean rules
        $\textsc{Rule}_k(\ell,n)$
\Ensure  integer array \texttt{classID[$i$]} $(i=1\ldots M)$
\For{$i \gets 1$ \textbf{to} $M$}
      \State $P \leftarrow P_i$
      \State $\ell \leftarrow$ \textsc{ToLabel}$(P)$
      \State $n \leftarrow$ \#\,non-identity symbols in $\ell$
      \For{$k \gets 0$ \textbf{to} $C-1$}
            \If{\textsc{Rule}$_k(\ell,n)$ \textbf{is true}}
                  \State \texttt{classID[$i$]} $\leftarrow k$
                  \State \textbf{break}
            \EndIf
      \EndFor
\EndFor
\State \Return \texttt{classID}
\end{algorithmic}
\end{algorithm}

After we classify all Pauli strings we are able to compute LOV weights and project them on the coefficient matrix. This projection scales the Pauli string coefficients of each Hamiltonian according to their geometric importance across the workload which is explicitly captured through local and global statistics and chemical structure used in the calculation of the LOV weights. Then using the projected matrix, LOV distances are calculated for adjacent geometries, i.e. for adjacent rows in the stacked matrix to capture adjacent Hamiltonian similarity, but also for the matrix, and thus the workload as a whole. This serves to quantify global maximum variation across geometries which serves as an upper limit for the dissimilarity two Hamiltonians could have across the whole workload. These procedures are explained in more detail below.

\paragraph{Computing LOV Weights and Weighted Projection.}
Utilizing the aligned Hamiltonian matrix \(\mathbf{C}\), we calculate the LOV weights for each Pauli class. These weights encode three statistical dimensions of importance: \emph{locality} (\(L\)) (Eq. \ref{eq:loc_def}), \emph{occupancy} (\(O\)) (Eq. \ref{eq:occ_def}), and \emph{variance} (\(V\)) (Eq. \ref{eq:var_def}).

Each component is computed via efficient vectorized operations across the full geometry workload. The final per-string LOV weight are computed and assembled in a diagonal matrix using Eq. \ref{eq:love_weight}.


To prepare for downstream similarity analysis, we project the aligned Hamiltonian coefficient matrix \(\mathbf{C}\) into a reweighted feature space:
\begin{equation}
  \mathbf{X} = \mathbf{C}\, W,
  \quad \text{where}\quad
  \mathbf{X} \in \mathbb{R}^{N \times P},
  \label{eq:weight_projection}
\end{equation}

This weighted projection amplifies the influence of more informative Pauli terms while suppressing low-impact variations, resulting in a distance metric that better reflects chemically relevant changes in the Hamiltonian.

\paragraph{LOV Distance Computation and Adaptive Thresholds.}
Using the LOV-weighted coefficient matrix \(\mathbf{X} = \mathbf{C} W\), we compute local pairwise distances between adjacent geometries and derive global variation thresholds for reuse and skipping.

For any fixed normalized state and adjacent Hamiltonians $H_g=H_{g-1}+\Delta H$, the Rayleigh quotient satisfies
\begin{equation}
  \left|\langle\psi|\Delta H|\psi\rangle\right|
  \leq \|\Delta H\|
  \leq \sum_{i=1}^{P}|\Delta c_i|
  \leq \sqrt{P}\|\Delta\mathbf{c}\|_2.
  \label{eq:hamiltonian_perturbation_bound}
\end{equation}
Thus, along smooth PES regions, small changes in aligned coefficients imply bounded energy changes. The positive LOV weighting $D_{\mathrm{LOV}}=\|\mathbf{w}\odot\Delta\mathbf{c}\|_2$ provides a structured surrogate that emphasizes chemically sensitive operator classes, while larger changes are routed to full optimization.

\paragraph*{Step 1 – Adjacent Distances.}
For consecutive geometries \(R_{g-1}, R_g\), the per-step LOV distance is
\begin{equation}
  D_{\mathrm{LOV}}(g) = \|x_{g} - x_{g-1}\|_{2},\quad g=1,\dots,N-1,
  \label{eq:lov_distance_single}
\end{equation}
where \(x_g\) is the \(g\)-th row of \(\mathbf{X}\). These values feed directly into the distance-aware scheduler used to calculate resource budgets for execution used by the multi-GPU engine.

\paragraph*{Step 2 – Global Maximum Variation.}
To establish adaptive thresholds, we compute each Pauli coefficient’s maximum deviation from the reference geometry \(R_0\):
\begin{equation}
  \Delta c_{\max,i} = \max_{g} |c_i(R_g) - c_i(R_0)|,
  \label{eq:delta_max_c}
\end{equation}
and aggregate them via LOV weights:
\begin{equation}
  D_{\max} = \|\mathbf{w} \odot \Delta c_{\max}\|_{2}
  = \sqrt{\sum_{i=1}^{P} (w_i\,\Delta c_{\max,i})^{2}}.
  \label{eq:d_max_scan}
\end{equation}

The maximum deviation of a coefficient is calculated by finding the coefficient in the same column, across all rows, of the stacked matrix with the biggest distance relative to the coefficient in the first row and calculating their difference. This process is repeated for each column and all maximum deviations are then scaled by their LOV weights and used to calculate $D_{max}$.

\paragraph*{Step 3 – Adaptive Thresholds.}
We define a surrogate Lipschitz constant representing the average LOV sensitivity:
\begin{equation}
  L_{\mathrm{avg}} = \|\mathbf{w}\|_{2}\sqrt{P}.
  \label{eq:l_avg}
\end{equation}
where $\mathbf{w}$ is the weight vector and $P$ is the number of Pauli terms of a single stacked Hamiltonian.

Given error tolerances \(\varepsilon_{\mathrm{skip}}\) and \(\varepsilon_{\mathrm{quick}}\), the scaling factor
\begin{equation}
  \eta_{\varepsilon}
    = \!\left(\frac{\varepsilon / L_{\mathrm{avg}}}{D_{\max} + \varepsilon / L_{\mathrm{avg}}}\right)^{1/3}
  \label{eq:eta_epsilon}
\end{equation}
produces dimensionless fractions that yield runtime thresholds
\(\varepsilon = \eta_{\varepsilon} D_{\max}\)
for reuse and skipping.  

The cubic root normalizes magnitude for stable scaling, while
\(L_{\mathrm{avg}}\propto \|\mathbf{w}\|_{2}\sqrt{P}\)
adapts thresholds to Hamiltonian size and sparsity,
enabling resource-aware scheduling aligned with system complexity.

We use energy tolerances $\varepsilon_{\mathrm{skip}}^{E}=1$~mHa and $\varepsilon_{\mathrm{quick}}^{E}=10$~mHa to derive the two adaptive fractions. The skip tolerance is deliberately stricter because skipped geometries receive no local optimization and reuse the preceding solution, consistent with thermochemical accuracy targets~\cite{curtiss1991gaussian2}. The quick tolerance is looser because these geometries still undergo local VQE refinement with reduced budgets, consistent with reaction-energy accuracy targets~\cite{reiher2017reaction}. The evaluation separately reports a $0.005$~Ha threshold to facilitate comparison with chemically relevant energy-error scales~\cite{bogojeski2020quantum,louvet2026feasibility}.

\subsubsection{Distance-Aware Budget Scheduler}
\label{subsubsec:distance_aware_scheduler}
Upon calculation of adjacent distances, the global maximum variation, and adaptive fractions, the distance-aware budget scheduler multiplies the fractions with the global maximum variation to calculate the thresholds for the quick and skip regime. The scheduler then uses the LOV distances to decide each geometry's regime and assign per-geometry budgets which are then passed to the multi-GPU pipeline for execution. This scheduling leverages these metrics to minimize overall resource utilization while maintaining accuracy. The exact procedures for assigning resource regimes and calculating the resource budget are defined and explained below.

\paragraph{Decision Bands and Thresholding.}
\label{subsubsec:decision_bands}

We first compute absolute thresholds $\varepsilon$ by scaling the maximum scan distance \(D_{\max}\) with the adaptive fractions $\eta$:
\begin{equation}
  \begin{aligned}
  \varepsilon_{\text{skip}} &= \eta_{\text{skip}} D_{\max}, &
  \varepsilon_{\text{quick}} &= \eta_{\text{quick}} D_{\max},\\
  0 &< \eta_{\text{skip}} < \eta_{\text{quick}} < 1,
  \end{aligned}
  \label{eq:thresholds}
\end{equation}

These thresholds establish decision bands to guide resource allocation for each geometry based on the computed LOV distance \(D_{\mathrm{LOV}}\):
\begin{itemize}[nosep,labelsep=0.5em]
\item \emph{Skip band} (\(D_{\mathrm{LOV}}\le\varepsilon_{\text{skip}}\)): Geometries within this band are sufficiently close to previously optimized points, so optimization is skipped entirely. The previously converged parameters and energies are reused (\(I_g=S_g=0\)).
\item \emph{Quick band} (\(\varepsilon_{\text{skip}}<D_{\mathrm{LOV}}\le\varepsilon_{\text{quick}}\)):
      Geometries moderately similar to prior points receive adaptive budgets.
      We scale iterations \(I_g\) and shots \(S_g\) inversely proportional to their
      normalized distance, allocating fewer resources to geometries closer to
      already optimized ones:
\begin{equation}
  I_{g}
    = \max\!\Bigl(1,\,
        \bigl\lceil \bigl(1 - D_{\mathrm{LOV}}/D_{\max}\bigr) c_{I} \bigr\rceil \Bigr),
  \label{eq:iteration_quick_band}
\end{equation}
\begin{equation}
  S_{g}
    = \max\!\Bigl(1,\,
        \bigl\lceil \bigl(1 - D_{\mathrm{LOV}}/D_{\max}\bigr) c_{S} \bigr\rceil \Bigr),
  \label{eq:shots_quick_band}
\end{equation}
      where the adaptive caps \(c_I\) and \(c_S\) are determined using the adaptive fractions $\eta$ (see Sec. \ref{subsubsec:similarity_analysis}) and the full budgets $I_{max}$ and $S_{max}$:
\begin{equation}
  c_{I} = \bigl\lfloor \eta_{\text{quick}} I_{\max} \bigr\rfloor,\quad
  c_{S} = \bigl\lfloor \eta_{\text{quick}} S_{\max} \bigr\rfloor,
  \label{eq:scalars_quick_band}
\end{equation}
\item \emph{Full band} (\(D_{\mathrm{LOV}}>\varepsilon_{\text{quick}}\)):
      Geometries significantly different from previous points receive the full
      user-defined computational budgets (\(I_g=I_{\max}\), \(S_g=S_{\max}\)).
\end{itemize}

The decision logic for allocating budgets is presented in Algorithm~\ref{algo:budget_decision}.

\begin{algorithm}[h!]
\caption{Distance-Aware Budget Decision}
\label{algo:budget_decision}
\small
\begin{algorithmic}[1]
\Require LOV distance $D_{\text{LOV}}$;
         thresholds $\varepsilon_{\text{skip}},\varepsilon_{\text{quick}}$;
         full budgets $(I_{\max},S_{\max})$;
         adaptive caps $(c_I,c_S)$
\Ensure  \texttt{skip} flag, optimizer iterations $I$, shot budget $S$
\If{$D_{\text{LOV}} \le \varepsilon_{\text{skip}}$}    \Comment{skip band}
      \State \Return\;(\textbf{true}, $0$, $0$)
\ElsIf{$D_{\text{LOV}} \le \varepsilon_{\text{quick}}$} \Comment{quick band}
      \State $I \leftarrow \max\left(1,\;\lceil(1 - \frac{D_{\text{LOV}}}{D_{\max}}) c_I\rceil\right)$
      \State $S \leftarrow \max\left(1,\;\lceil(1 - \frac{D_{\text{LOV}}}{D_{\max}}) c_S\rceil\right)$
      \State \Return\;(\textbf{false}, $I$, $S$)
\Else                                             \Comment{full band}
      \State \Return\;(\textbf{false}, $I_{\max}, S_{\max}$)
\EndIf
\end{algorithmic}
\end{algorithm}

\smallskip
\paragraph{Budget Assignment Strategy.}
Due to parameter-transfer warm-starting, the first geometry must be solved from a classical initialization (e.g., Hartree-Fock state \cite{hf_2020}). This initial optimization provides the reference parameters \(\theta_0\) and energy \(E_0\) necessary to bootstrap subsequent geometries.

With the LOV distances and decision bands computed offline, the distance-aware scheduler precomputes and stores a resource budget table containing iteration and shot allocations for each geometry. These budgets are determined and fixed \emph{prior} to any quantum circuit execution. Consequently, the \sol~framework expends zero quantum or classical resources on geometries classified within the skip band and assigns iteration and shot budgets to quick band geometries identified as near previously converged points.

By allocating resources through offline analysis, the scheduler efficiently distributes computational effort across the workload, reducing runtime while preserving chemical accuracy.  
Budget over- or under-estimation arises from the trade-off between the metric’s conservatism—weights emphasizing geometry-sensitive operator classes—and its responsiveness to small perturbations.  
Over-estimation simply allocates extra effort near regions of strong structural change, maintaining physical robustness, while under-estimation occurs only when all operator classes vary weakly, where the problem is already smooth and inexpensive to solve.

\subsubsection{Multi-GPU VQE Engine}
\label{subsubsec:multi_gpu_vqe}

\paragraph{Batched Gradient Dispatch.}

With Hamiltonian similarity analysis completed (Sec.~\ref{subsubsec:similarity_analysis}) and geometry-specific budgets assigned (Sec.~\ref{subsubsec:distance_aware_scheduler}), the Multi-GPU VQE engine executes the full PES workload via batched gradient computations across available GPUs.

\paragraph{{Persistent CUDAQ Solver with Deep Warm-Start}}
\label{subsubsec:cudaq_solver}

To reduce per-geometry optimization cost, we design a persistent CUDAQ solver that integrates parameter transfer, optimizer state caching, and incremental Hamiltonian updates. For the initial geometry $R_0$, a classical Hartree–Fock initialization is used \cite{hf_2020}. The resulting converged parameters \(\theta_0\) are cached and reused to warm-start the optimization of $R_1$, with subsequent geometries warm-started from their immediate predecessors (see Sec.~\ref{subsec:warm_start}). This amortizes the cost of classical initialization over the full scan.

The solver constructs a long-lived CUDAQ context at molecule load time and compiles the variational ansatz into a persistent CUDA graph. This graph is replayed at each geometry step, avoiding repeated compilation and launch overheads. Importantly, we cache not only \(\theta\) but also the full optimizer state (e.g., gradients, momentum buffers, quasi-Newton terms). This “deep” warm-start accelerates convergence by preserving momentum across geometry steps.

To reduce GPU communication overhead, we introduce an \emph{Incremental Hamiltonian} structure that streams only coefficient differences between adjacent geometries. This exploits the column alignment performed during similarity analysis and ensures minimal GPU data movement during solver updates.

%% file: sections/evaluation-new.tex
\section{Evaluation}
\label{sec:evaluation}

We evaluate \textsc{OptiPESS} on molecular PES workloads and focus on three questions:
(i) how much end-to-end runtime and quantum-resource reduction LOV-based scheduling provides,
(ii) whether adaptive budgeting preserves VQE accuracy, and
(iii) whether offline LOV-based similarity computation incurs significant overhead.

\begin{figure*}[t]
    \centering
    \includegraphics[width=0.95\textwidth]{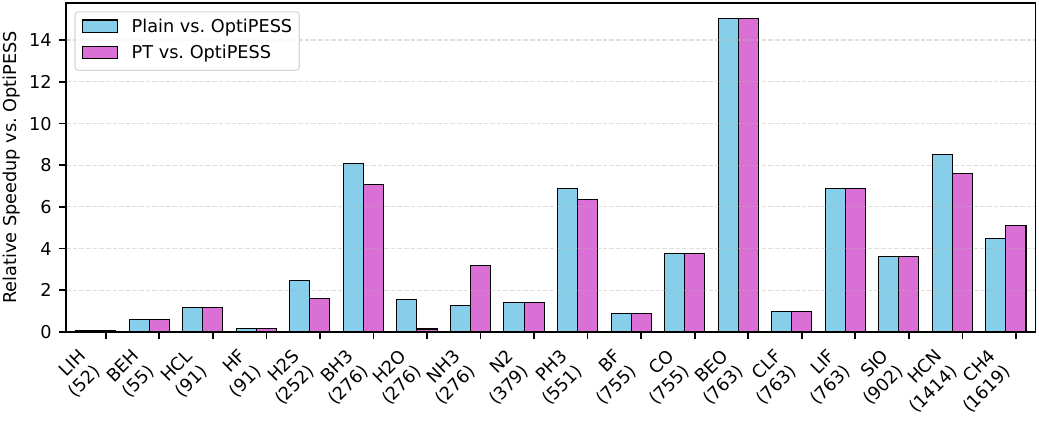}
    \caption{Evaluation speedup of \sol~relative to Plain and PT across the PES dataset (noiseless CUDA-Q simulation).}
    \label{fig:overall_speedup}
\end{figure*}

\subsection{Experimental Setup}

\paragraph*{Implementation and Environment.}
The implementation follows the pipeline in Sec.~\ref{subsec:pipeline}:
\begin{enumerate}[leftmargin=1.5em,label=\arabic*),itemsep=1pt,topsep=2pt,parsep=0pt]
\item \emph{Molecular problem generation:} Qiskit Nature~0.7.3 and PySCF generate molecular problems and fermionic operators. OpenFermion~1.7 and OpenFermion-PySCF~0.5 convert them to qubit Hamiltonians.
\item \emph{Hamiltonian alignment:} The generated Hamiltonians are aligned into a common Pauli-label basis for LOV analysis and budget construction (Sec.~\ref{subsec:ham_align}).
\item \emph{LOV scoring and budgeting:} CuPy performs GPU-resident alignment, LOV scoring, Hamiltonian-distance calculations, and budget construction (Secs.~\ref{subsec:gpu_similarity} and~\ref{subsubsec:distance_aware_scheduler}).
\item \emph{CUDA-Q execution:} CUDA-Q~0.11 with CUDA~12 and Python~3.11 executes UCCSD VQE through the NVIDIA \texttt{nvidia} target with the \texttt{mqpu} option~\cite{CUDA-Q}. Each GPU is exposed as a simulated QPU backed by cuStateVec, enabling asynchronous parallel execution (Sec.~\ref{subsec:multi_gpu_runtime}).
\item \emph{Classical optimization and analysis:} NumPy~1.26.4 and SciPy support CPU-side processing, SLSQP optimization, and result aggregation.
\end{enumerate}
All end-to-end results in this section are obtained from \emph{noiseless CUDA-Q statevector simulations} on eight NVIDIA A100 GPUs (SaturnCloud) and 32 Intel Xeon Platinum 8358 CPUs, with distributed parameter-shift gradients evaluated asynchronously and a persistent CUDA-Q context.

\paragraph*{Workloads.}
We use PES scans for 18 molecules spanning 6–14 qubits and 52–1619 Pauli terms. Table \ref{tab:molecule_summary_sorted_vertical_columns} lists these molecules in increasing order of number of qubits and includes the number of Paulis for each molecule. The average time for one circuit evaluation for each molecule is listed in Table \ref{tab:compact_optipess}.

\begin{table}[ht]
\caption{Benchmark Molecules and Pauli-Term Counts}
\label{tab:molecule_summary_sorted_vertical_columns}
\centering
\resizebox{0.85\linewidth}{!}{%
\begin{tabular}{|lll|lll|}
\hline
\textbf{Mol} & \textbf{Qubits} & \textbf{Paulis} & \textbf{Mol} & \textbf{Qubits} & \textbf{Paulis} \\
\hline
LIH  & 6  &   52   & PH3  & 12 &  551   \\
BEH  & 8  &   55   & BF   & 12 &  755   \\
HF   & 10 &   91   & CO   & 12 &  755   \\
HCL  & 10 &   91   & BeO  & 12 &  763   \\
H2S  & 10 &  252   & CLF  & 14 &  763   \\
BH3  & 10 &  276   & LIF  & 12 &  763   \\
H2O  & 10 &  276   & SIO  & 14 &  902   \\
NH3  & 10 &  276   & HCN  & 14 & 1414   \\
N2   & 12 &  379   & CH4  & 12 & 1619   \\
\hline
\end{tabular}%
}
\end{table}

\paragraph*{Baselines and Metrics.}
We compare \sol~against:
Plain VQE without warm-start or adaptive budgeting, and
PT (parameter transfer with warm-start but fixed budgets).
All methods share the same UCCSD ansatz, SLSQP optimizer, iteration cap $I_{\max}=50$, and shot cap $S_{\max}=1024$.
We report the total number of optimizer iterations, total circuit evaluations (shots), end-to-end CUDA-Q wall time, mean percent energy deviation vs.\ each baseline, and average absolute error vs.\ exact diagonalization.  
The evaluation accuracy threshold is 0.005~Ha, with the scheduler tolerances specified in Sec.~\ref{subsubsec:similarity_analysis}.
We define speedup as 
$$  \text{Speedup}(R)
    \;=\;
    \frac{N_{\text{base}}(R)}{N_{\text{OPTIPESS}}(R)} - 1
$$
as in \cite{Skogh_2023} for consistency.
We note that the terms circuit evaluations and shots are used interchangeably as shots represent the times a quantum circuit is evaluated on real hardware in order to extract the  result from statistical fluctuations. We use circuit evaluations in the rest of the evaluation section for clarity to less familiar readers and to emphasize the savings \sol~achieves. Additionally, we make the distinction that noiseless simulations are exact analytic solutions and represent one circuit evaluation while noisy simulations are run with shots set to 1024 which equates to 1024 executions of the quantum circuit for each expectation evaluation.

\subsection{End-to-End Performance and Accuracy (CUDA-Q Simulation)}

\paragraph*{Performance in Noiseless Simulations.} Figure~\ref{fig:overall_speedup} summarizes dataset-wide reductions in evaluations and CUDA-Q simulation time. Averaged over the full dataset, we observe a $76.4\%$ reduction in evaluations and a $70.8\%$ reduction (up to $74.8\%$ in Table \ref{tab:method_metrics}) in wall time relative to Plain, and comparable reductions relative to PT, while the mean energy deviation remains below $0.03\%$ for both comparisons. Therefore, runs evaluate significantly fewer circuits and iterations without degrading the converged energies.

The benefit increases with Hamiltonian size.  
For molecules with fewer than 200 Pauli terms, LOV-based scheduling yields consistent but modest gains (about $1.4$–$1.8\times$ end-to-end speedup).
For Hamiltonians with more than 1000 Pauli terms, end-to-end speedups reach approximately $3.0$–$3.6\times$.
In these larger systems, the Pauli landscape is denser and more anisotropic. The LOV embedding separates “smooth” regions of the PES (where parameter reuse is safe) from highly varying regions, enabling more aggressive budget compression than simple Frobenius or HS-distance heuristics while maintaining convergence as shown in Table \ref{tab:method_metrics}.

\begin{table}[h!]
\caption{Performance Across Similarity Measures in Noiseless Simulation}
\label{tab:method_metrics}
\centering
\resizebox{1.0\columnwidth}{!}{%
\begin{tabular}{lccc}
\toprule
\textbf{Method} & \textbf{Evals.} & \textbf{Time (s)} & \textbf{\% Diff (PT/Plain)} \\
\midrule
Plain            & 8674 & 11908.82 & N/A \\
PT               & 9976 & 11013.74 & N/A \\
\sol+Frobenius        & 3092 & 3657.43  &  -0.0338 / -0.0412 \\
\sol+Hilbert-Schmidt  & 5079 & 6477.84  &  +0.0671 / +0.0596\\
\sol+LOV             & 2371 & 3002.23  &  -0.0143 / -0.0218 \\
\bottomrule
\end{tabular}
}
\end{table}

\paragraph*{Performance in Noisy Simulations.}
To study the impact of adaptive budgets with noise, Table~\ref{tab:shot_metrics} reports the total number of circuit evaluations in noisy runs, contrasting fixed and adaptive quick-band iteration caps.

\begin{table}[h!]
\caption{Noisy-Run Circuit Evaluations and Average Error Difference Across All Molecules}
\label{tab:shot_metrics}
\centering
\resizebox{1.0\columnwidth}{!}{%
\begin{tabular}{lrr}
\toprule
Method & Circ. Evals & \% error diff (PT / Plain) \\
\midrule
Plain                      & 4{,}764{,}672 & N/A \\
PT                         & 5{,}121{,}024 & N/A \\
\sol~(fixed quick iters)   & 1{,}319{,}870 & $-0.003$ / $-0.0007$ \\
\sol~(adaptive quick iters)& 1{,}109{,}790 & $-0.0288$ / $-0.0264$ \\
\bottomrule

\end{tabular}
}
\end{table}

With fixed quick-band iterations, \sol{} reduces circuit evaluations by  
3.44M (72.3\%) relative to Plain and 3.80M (74.2\%) relative to PT, while achieving essentially identical energies.
Enabling adaptive quick-band iterations provides even larger savings—3.65M (76.7\%) and 4.01M (78.3\%) fewer evaluations relative to Plain and PT—while slightly improving average error.
Since each optimizer step requires at least one energy evaluation, these reduced circuit evaluations counts correspond directly to fewer quantum-circuit executions, the dominant cost on real hardware.

\paragraph*{Performance on Different Optimizers.} Testing \sol~on a small dataset (molecules with fewer than 500 Paulis) with L-BFGS-B and SPSA confirms its effectiveness in reducing iterations while increasing accuracy, as shown in Table \ref{tab:optimizer_combos_noevals}. The accuracy improvement of SPSA relative to L-BFGS-B and SLSQP follows the trend observed in \cite{kanbench2024}.

\begin{table}[h!]
\caption{\sol~Performance with L-BFGS-B and SPSA}
\label{tab:optimizer_combos_noevals}
\centering
\small
\setlength{\tabcolsep}{4pt}
\begin{tabular}{llrrr}
\toprule
\textbf{Opt.} & \textbf{Mode} & \textbf{Iters.} & \textbf{Time (s)} & \textbf{\% Diff (PT/Plain)} \\
\midrule
\multirow{3}{*}{L-BFGS-B}
  & Plain & 2{,}344  & 38{,}999.93 & -- \\
  & PT    & 1{,}562  & 29{,}597.45 & -- \\
  & \sol  & 522      & 9{,}021.30  & $-3.64 / -5.02$ \\
\midrule
\multirow{3}{*}{SPSA}
  & Plain & 64{,}800 & 6{,}700.16  & -- \\
  & PT    & 34{,}200 & 3{,}557.63  & -- \\
  & \sol  & 29{,}656 & 3{,}193.99  & $-48.7 / -4.8$ \\
\bottomrule
\end{tabular}
\end{table}

\begin{figure*}[t]
    \centering
    \includegraphics[width=0.95\textwidth]{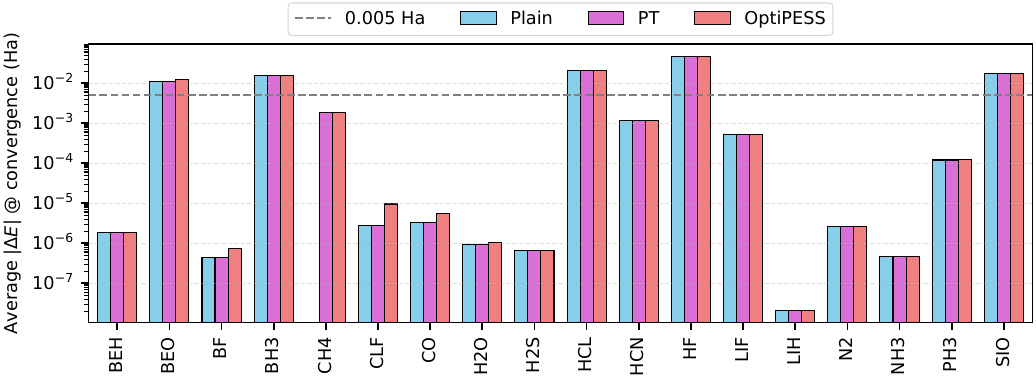}
    \caption{Average absolute error at the 3 lowest-energy geometries per molecule (evaluation threshold = 0.005~Ha).}
    \label{fig:lowest_energy_errors}
\end{figure*}

\paragraph*{Accuracy Insights.} Accuracy at chemically relevant geometries is shown in Fig.~\ref{fig:lowest_energy_errors}.
For the three lowest-energy geometries per molecule, \sol~achieves errors below $0.005$~Ha for the vast majority of molecules and matches or exceeds the static baselines where deviations occur.  
If we stop the Plain baseline at the iteration where \sol~has already converged, Plain exhibits higher residual error in 7 of 10 representative molecules, indicating that the LOV-based stopping decisions track convergence at least as well as, and often better than, a fixed-budget schedule.

\subsection{Quantum Cost Savings}

Finally, we quantify savings in terms of single-GPU CUDA-Q simulation time.  
Table~\ref{tab:compact_optipess} reports, for each molecule, the average time to simulate one circuit evaluation on a single A100 ($T_Q$) over 10 repetitions, the reduction in optimizer iterations versus Plain, and the corresponding reduction in CUDA-Q simulation time.
Since every optimizer iteration requires at least one circuit evaluation (and typically many for gradient estimation), reduced iterations translate directly into fewer quantum circuit executions on a real device.

\begin{table}[h!]
\caption{Quantum Savings Under CUDA-Q Simulation}
\label{tab:compact_optipess}
\centering
\setlength{\tabcolsep}{3pt}
\renewcommand{\arraystretch}{1.08}
\footnotesize
\begin{tabular}{@{}lrrr@{}}
\hline
\textbf{Mol} & \textbf{$T_Q$ (s)}
& \textbf{$\Delta I$ (Plain/PT)}
& \textbf{$\Delta T_Q$ (s, Plain/PT)}\\
\hline
beh2 & 0.979  & (95 / 93)       & (93.0 / 91.0) \\
h2o  & 1.200  & (331 / -246)    & (397.2 / -295.2) \\
hf   & 0.969  & (30 / 30)       & (29.1 / 29.1) \\
lih  & 0.071  & (10 / 10)       & (0.71 / 0.71) \\
n2   & 2.577  & (239 / 239)     & (616.9 / 616.9) \\
nh3  & 1.153  & (216 / 548)     & (249.1 / 631.8) \\
hcl  & 1.056  & (245 / 245)     & (258.7 / 258.7) \\
h2s  & 1.114  & (336 / 219)     & (374.3 / 244.0) \\
bh3  & 1.750  & (1421 / 1241)   & (2486.8 / 2171.8) \\
beo  & 5.188  & (2179 / 2179)   & (11307.6 / 11307.6) \\
lif  & 5.188  & (1515 / 1515)   & (7859.8 / 7859.8) \\
ph3  & 3.747  & (1080 / 998)    & (4046.8 / 3738.3) \\
co   & 5.134  & (771 / 771)     & (3956.3 / 3956.3) \\
bf   & 5.134  & (147 / 147)     & (754.7 / 754.7) \\
clf  & 6.053  & (221 / 221)     & (1338.7 / 1338.7) \\
sio  & 7.155  & (1418 / 1418)   & (10146.7 / 10146.7) \\
hcn  & 11.217 & (2022 / 1800)   & (22698.8 / 20190.6) \\
ch4  & 11.008 & (626 / 714)     & (6890.0 / 7861.7) \\
\hline
Total
& -- 
& (12902 / 12142) 
& (73484.2 / 70898.6) \\
\hline
\end{tabular}
\end{table}

Over the full dataset, \sol~eliminates about 12.9K optimizer iterations relative to Plain, corresponding to more than $7\times10^{4}$ seconds of single-GPU CUDA-Q simulation time.
For larger molecules such as SiO and HCN, the savings per PES scan reach several hours of avoided simulation time, which also directly translate to substantial reductions in QPU usage on real hardware.
These results highlight that LOV-guided scheduling is not only an algorithmic optimization but also a practical lever for reducing expensive quantum circuit evaluations in hybrid quantum–classical pipelines.

\subsection{GPU Scaling and Static-Phase Overhead}

\begin{figure}[h!]
    \centering
    \includegraphics[width=\linewidth]{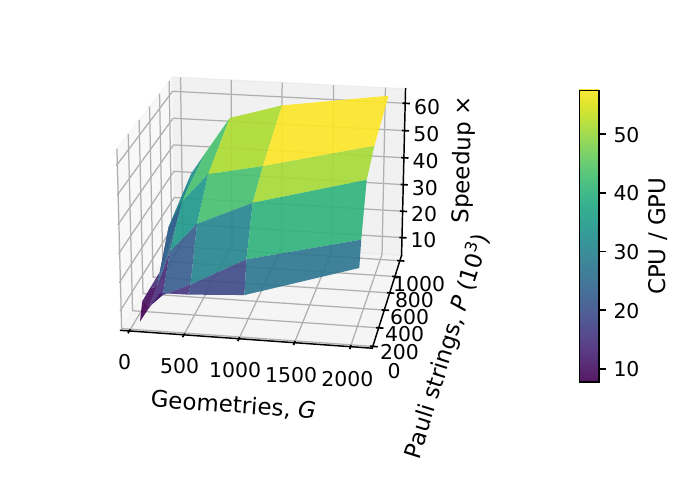}
    \caption{GPU speedup for LOV analysis relative to CPU. Increasing the number of geometries by $32\times$ and Pauli strings by $20\times$ yields over $50\times$ GPU speedup.
    }
    \label{plot:lov_speedup_cpu_gpu}
\end{figure}

The static LOV phase runs entirely on GPU and can be reused across different runtime configurations (optimizers, ansätze, or shot schedules).  
Table~\ref{tab:lov_bench_params} lists the benchmark grid used to stress this phase, and Fig.~\ref{plot:lov_speedup_cpu_gpu} reports speedups over a NumPy-based CPU implementation as a function of geometries and Pauli strings.

\begin{table}[h!]
  \caption{CPU and GPU LOV Benchmark Grid}
  \label{tab:lov_bench_params}
  \centering
  \renewcommand{\arraystretch}{1.00}
  \begin{tabular}{@{}l l@{}}
    \toprule
    Parameter & Values \\ \midrule
    Geometries $G$      & $64, 128, 256, 512, 1024, 2048$ \\
    Pauli strings $P$   & $50{\times}10^{3}, 100{\times}10^{3}, 250{\times}10^{3}, 500{\times}10^{3}, 10^{6}$ \\
    \bottomrule
  \end{tabular}
\end{table}

As both $G$ and $P$ grow, the CuPy implementation scales strongly, moving from about $4.5\times$ speedup at $(G{=}64,P{=}50\mathrm{K})$ to roughly $64\times$ at $(G{=}2048,P{=}10^6)$. The molecular workload contains 324 geometry instances and, summing Table~\ref{tab:molecule_summary_sorted_vertical_columns}, 10,033 Pauli terms per geometry set (180,594 geometry--Pauli instances). Static analysis remains below $2\%$ of the 3002.23~s end-to-end \sol+LOV time. For comparison, the much larger $(G{=}2048,P{=}10^6)$ benchmark takes only $0.0857+0.0569+0.0310=0.174$~s on one A100 for LOV weights, adjacent distances, and $D_{\max}$ ($0.006\%$ of 3002.23~s). Its CPU time is 11.12~s ($0.37\%$).
The analytical runtime and memory scaling of this preprocessing stage are derived in Sec.~\ref{subsec:lov_cost_analysis}.


%% file: sections/discussion-future-work.tex
\section{Discussion and Future Work}
\label{sec:discussion}

\sol~is most directly applicable to ordered families of related Hamiltonians, such as PES scans, where adjacent instances usually change smoothly. Smooth regions produce more skip/quick decisions, while abrupt changes trigger full optimization and reduce savings. The relevant workload property is therefore cross-instance Hamiltonian continuity rather than molecule size alone.

Because LOV is defined at the Hamiltonian-coefficient level, it is not tied to a particular variational ansatz. We confirmed implementation compatibility on QAOA-style Hamiltonian batches, while the quantitative evaluation focuses on VQE because it is the domain-relevant algorithm for molecular ground-state PES scans. For example, in weighted Max-Cut
each instance corresponds to a cost Hamiltonian $H_C = \sum_{(i,j)} w_{ij} Z_i Z_j$. 
By grouping Pauli terms according to graph topology (edge degree, community structure, sparsity), \sol~can allocate budgets adaptively across graph instances just as it does across molecular geometries.

The benchmark set uses small and medium active spaces so exact references and repeated controlled comparisons remain practical. Larger molecular simulations require active spaces, qubit counts, and statevector memory beyond the available allocation. The synthetic study therefore isolates LOV scaling up to one million Pauli terms. \sol~demonstrated strong scaling on multi-GPU nodes, but extending to multi-node settings introduces interconnect and synchronization costs.
Future work includes distributed execution across heterogeneous GPU clusters and hybrid QPU backends.  
A natural extension is closed-loop scheduling: adjusting iteration and circuit evaluation budgets online based on observed variance or hardware feedback rather than relying solely on offline analysis.


%% file: sections/conclusion-new.tex
\section{Conclusion}
\label{sec:conclusion}

We presented \textsc{OptiPESS}, a runtime system for adaptive resource allocation in hybrid quantum–classical workloads.
By introducing the Locality–Occupancy–Variance (LOV) metric and combining it with a distance-aware scheduler and multi-GPU CUDA-Q backend, OptiPESS transforms static variational algorithms into predictive, resource-efficient executions.
The system reduces both classical and quantum computation on PES Scans on a dataset of 18 molecules by up to 3.6× in noiseless simulations and reduces computation by up to $78.3\%$ in noisy simulations while maintaining or improving chemical-accuracy-level precision across workloads. The LOV similarity showed strong scaling as the problem sizes got many orders of magnitude larger while accounting for less than $2\%$ of the total wall time. 

OptiPESS demonstrates that lightweight offline analysis and deterministic runtime scheduling can substantially improve the efficiency of hybrid quantum computation without modifying underlying algorithms.
This architectural co-design approach bridges the gap between algorithmic theory and practical system implementation, paving the way for scalable, cost-aware execution of large Hamiltonian workloads on next-generation quantum–classical platforms.
